\documentclass[11pt]{article}
\usepackage{moriond}

\def\Journal#1#2#3#4{{#1} {\bf #2}, #3 (#4)}

\def\PRL{\em Phys. Rev. Lett.}
\def\PRD{{\em Phys. Rev.} D}

\def\EPJC{{\em Eur.Phys.J.} C}

\def\be{\begin{equation}}
\def\ee{\end{equation}}
\def\bea{\begin{eqnarray}}
\def\eea{\end{eqnarray}}

\def\be{\begin{equation}}
\def\ee{\end{equation}}

\def\bi{\begin{itemize}}
\def\ei{\end{itemize}}
\def\bc{\begin{center}}
\def\ec{\end{center}}
\def\and{\/\mbox{and}}

\newcommand{\squishlist}{
 \begin{list}{$\bullet$}
  { \setlength{\itemsep}{0pt}
     \setlength{\parsep}{3pt}
     \setlength{\topsep}{3pt}
     \setlength{\partopsep}{0pt}
     \setlength{\leftmargin}{1.5em}
     \setlength{\labelwidth}{1em}
     \setlength{\labelsep}{0.5em} } }

\newcommand{\squishlisttwo}{
 \begin{list}{$\bullet$}
  { \setlength{\itemsep}{0pt}
     \setlength{\parsep}{0pt}
    \setlength{\topsep}{0pt}
    \setlength{\partopsep}{0pt}
    \setlength{\leftmargin}{2em}
    \setlength{\labelwidth}{1.5em}
    \setlength{\labelsep}{0.5em} } }

\newcommand{\squishend}{
  \end{list}  }

\newcommand{\TeV}{\ensuremath{\mathrm{Te\kern -0.1em V}}}
\newcommand{\GeV}{\ensuremath{\mathrm{Ge\kern -0.1em V}}}
\newcommand{\MeV}{\ensuremath{\mathrm{Me\kern -0.1em V}}}

\newcommand{\MeVCSq}{\ensuremath{\MeV\!/c^2}}
\newcommand{\GeVCSq}{\ensuremath{\GeV\!/c^2}}

\newcommand{\jpsi}{\ensuremath{J/\psi}}

\newcommand{\pbinv}{\ensuremath{\,\mathrm{pb}^{-1}}}
\newcommand{\fbinv}{\ensuremath{\,\mathrm{fb}^{-1}}}

\newcommand{\met}{\mbox{${\hbox{$E$\kern-0.6em\lower-.1ex\hbox{/}}}_{\rm T}\:$}}
\newcommand{\metx}{\mbox{${\hbox{$E$\kern-0.6em\lower-.1ex\hbox{/}}}_T\:$}}
\newcommand{\mety}{\mbox{${\hbox{$E$\kern-0.6em\lower-.1ex\hbox{/}}}_T\:$}}

\newcommand{\bit}{\begin{itemize}}
\newcommand{\eit}{\end{itemize}}
\newcommand{\bce}{\begin{center}}
\newcommand{\beqn}{\begin{eqnarray*}}
\newcommand{\eeqn}{\end{eqnarray*}}
\newcommand{\beq}{\begin{equation}}
\newcommand{\ece}{\end{center}}

\newcommand{\Photo}{\vspace{0.5cm}\includegraphics[height=35mm]{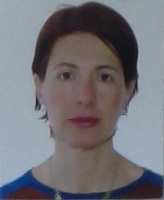}}

\begin{document}
\vspace*{4cm}
\title{B AND C SPECTROSCOPY AT LHCB}

\author{ G.~MANCA }

\address{Department of Physics, Universit\`a degli studi di Cagliari, \\ 
S.P. per Sestu, Km 0.7, 09042 Monserrato, Cagliari, Italy}

\maketitle\abstracts{
We will present recent results in the field of b and c spectroscopy 
at LHCb, with particular attention to the latest studies on the X(3872) 
quantum numbers and the $B_c$ new decay modes and mass measurement.
}

\section{Introduction}
The production of mesons and baryons containing
$b$ and $c$ quarks is copious at the LHC.
The studies on the production and spectroscopy of these
particles are important inputs to other measurements 
and bring valuable contributions to the thorough 
understanding of the mechanisms of QCD production.
The LHCb detector~\cite{lhcb} has a unique geometry 
optimised for these studies, as it accepts 40\%
of all $B$ hadrons produced in $pp$ interactions.
The detector is a single-arm forward spectrometer~\cite{lhcb} 
dedicated to flavour physics at the LHC. 
In the years from 2010 to 2012  LHCb has recorded an integrated luminosity of
about 3~\fbinv of data at a center-of-mass energy of 
7, 2.76 and 8 TeV with an efficiency of more than 90\%.
We discuss recent $B_c$ and b-baryons results in Section~\ref{sec:bc} and 
~\ref{sec:bmasses}, 
and the X(3872) results in Section~\ref{sec:xjpc}.

\section{$B_c$ physics}\label{sec:bc}
The $B_c$ is  the only B meson 
made of two ``heavy'' quarks and, as such, its properties 
are in between the charmonium and bottomonium states.
It was first observed by CDF in 1998~\cite{bc1} in the $J/\psi \ell^\pm\nu$
decay mode, and fully reconstructed the  $J/\psi \pi^\pm$
mode~\cite{bc2}. At LHCb we already measured the $B_c$ mass and production 
cross section in the latter channel using ~40~\pbinv
of data~\cite{bc3}. In the larger dataset  we 
observed two new decay modes, $B_c^\pm\to \psi(2S) \pi^\pm$
and $B_c^\pm\to J/\psi D_s^{(*)\pm}$.
\subsection{$B_c^\pm\to \psi(2S) \pi^\pm$ observation}
\begin{figure}
\begin{minipage}{0.345\linewidth}
\centerline{\includegraphics[width=\linewidth]{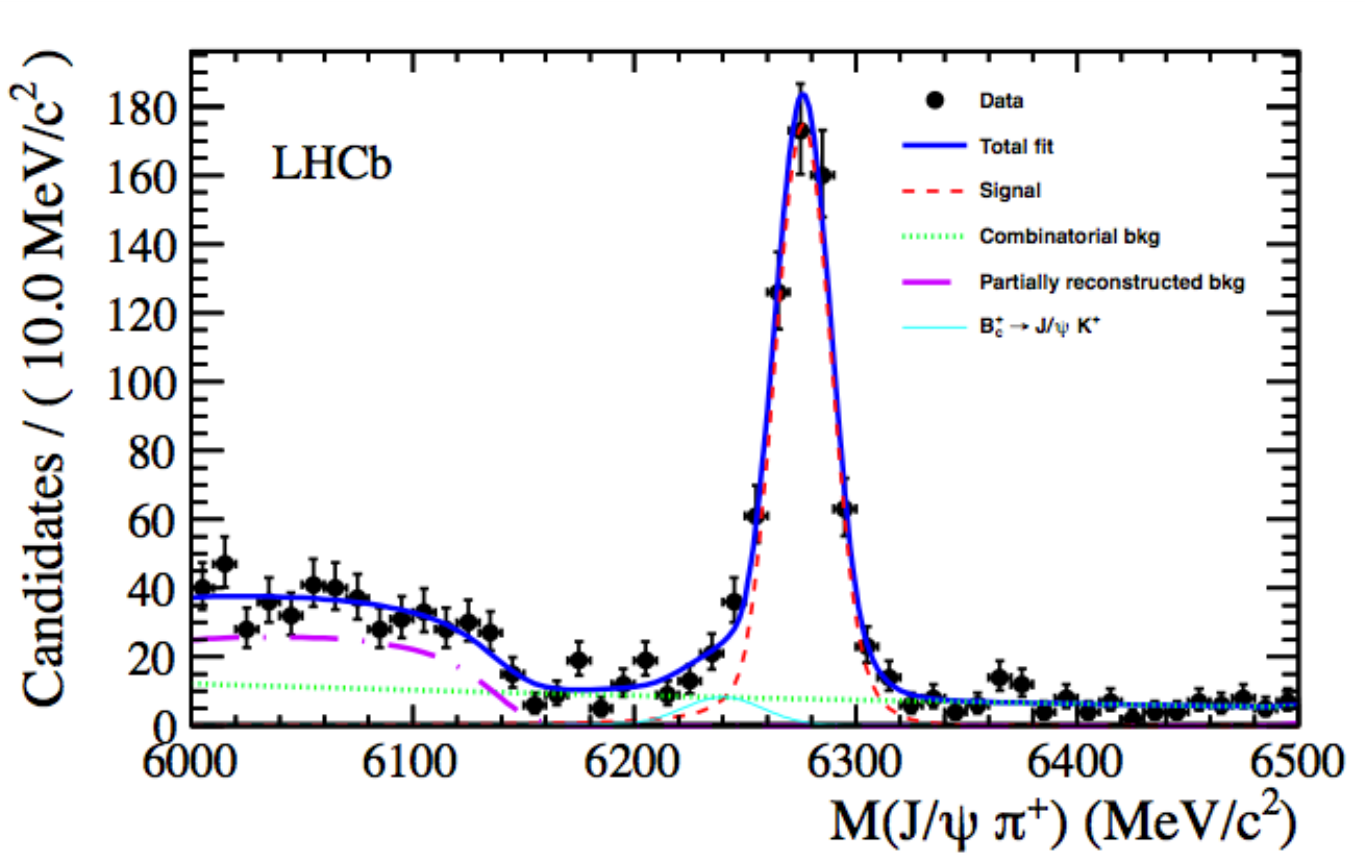}}
\end{minipage}
\hfill
\begin{minipage}{0.345\linewidth}
\centerline{\includegraphics[width=\linewidth]{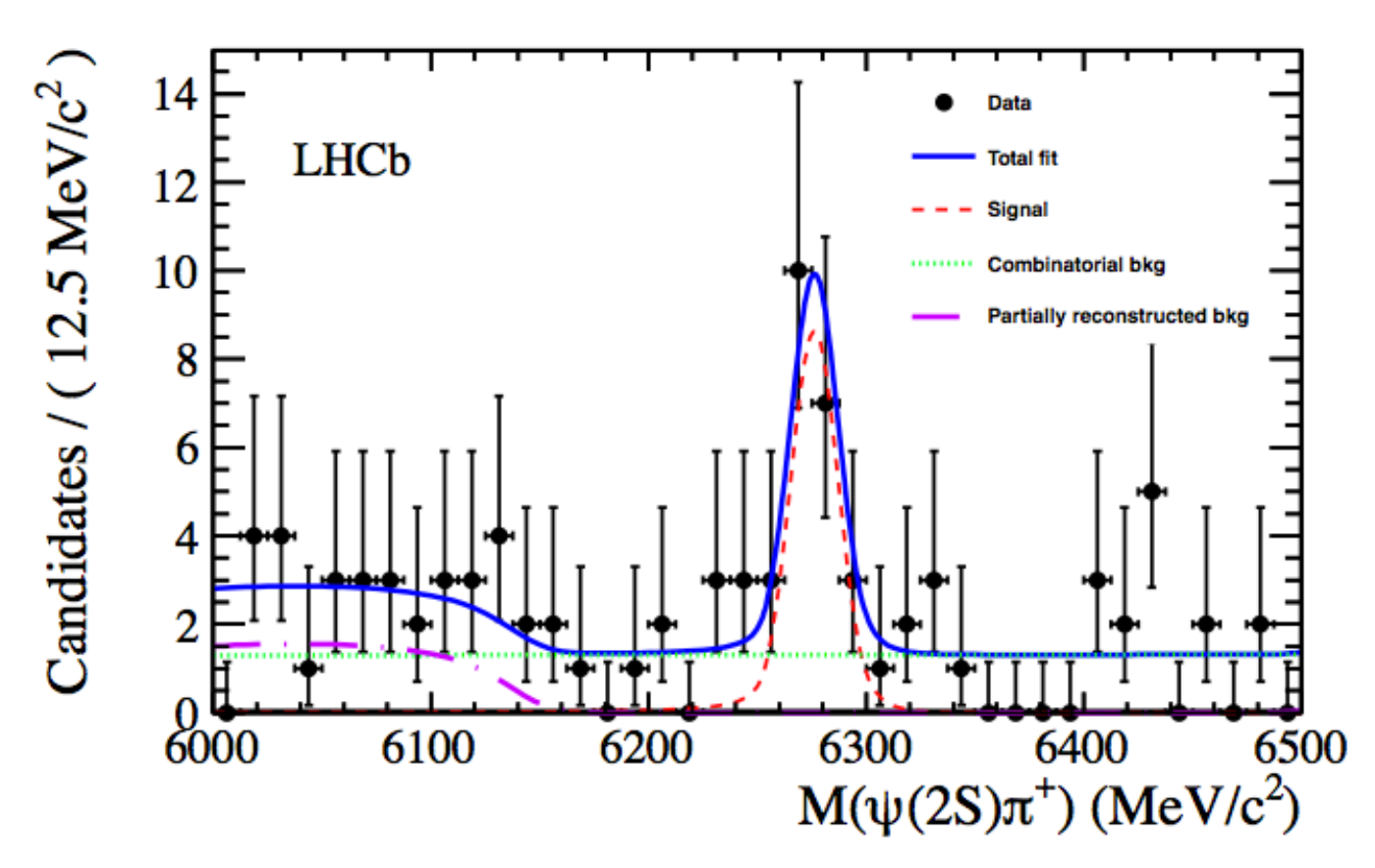}}
\end{minipage}
\hfill
\begin{minipage}{0.29\linewidth}
\centerline{\includegraphics[width=\linewidth]{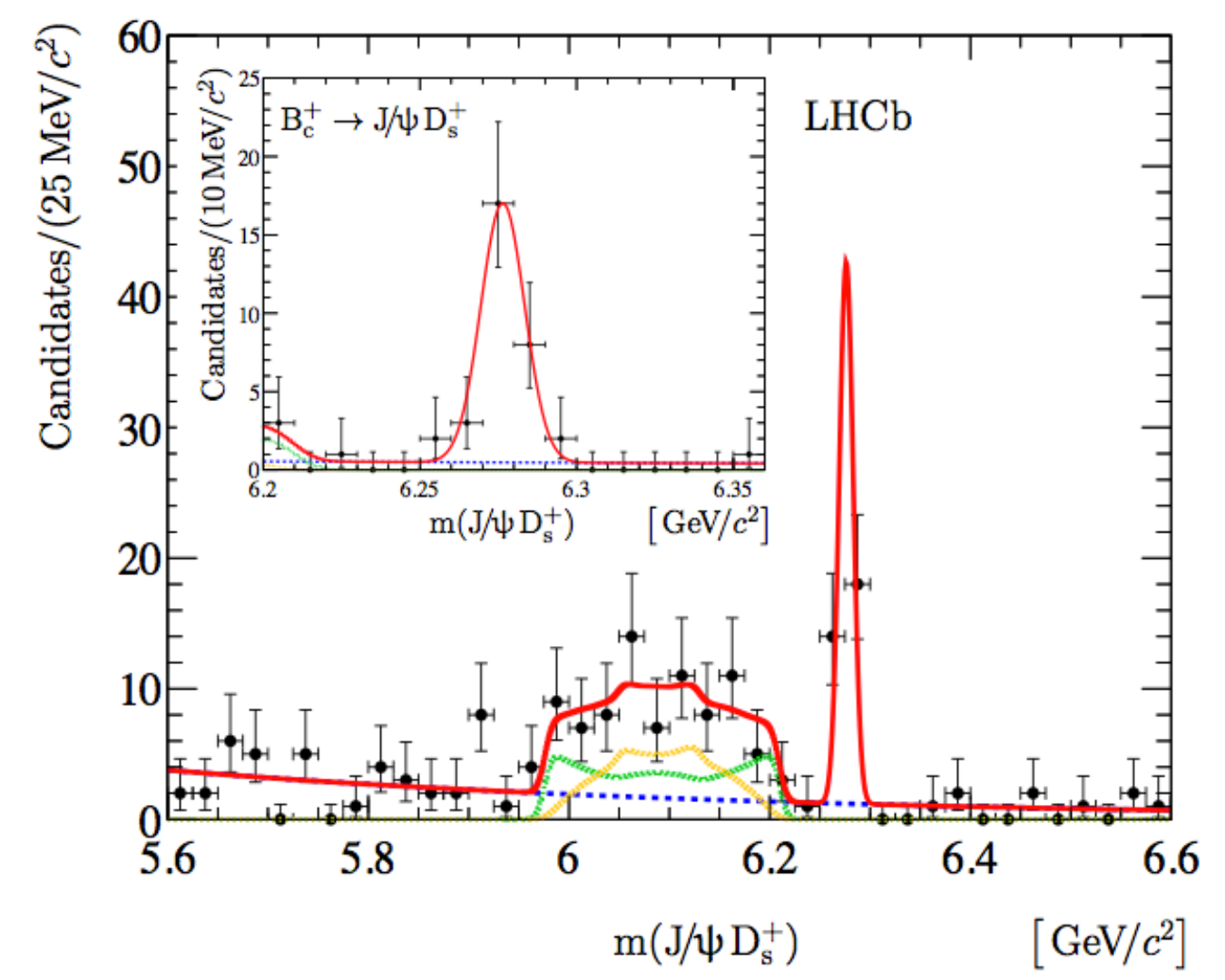}}
\end{minipage}
\caption{$\jpsi \pi^\pm$(left), $\psi(2S) \pi^\pm$(middle) and $D_s^{\pm(*)} \pi^\pm$(left) 
invariant mass distributions. The inset in the leftmost figure is the fit in the region
between 6.2 and 6.35~\GeVCSq performed with different binning.}
\label{fig:Fig1}
\end{figure}

In 1.1~\fbinv of data we observed
595$\pm$29  $B_c \to J/\psi \pi^\pm$ 
and  20$\pm$5 $B_c \to \psi(2S) \pi^\pm$ 
candidate events (as shown in Fig.~\ref{fig:Fig1}, left and middle), 
selected  using the 
Boost Decision Tree (BDT) technique~\cite{bdt} trained on the 
$B_c \to J/\psi \pi^\pm$ more abundant 
channel. 
The number of signal candidates
is determined by fitting the $\psi(2S) \pi$ 
invariant mass distribution with a Crystal Ball function~\cite{cb} for the signal and 
an exponential function for the background.
Partially reconstructed events and combinatorial background
are also accounted for.
We measured the ratio of  $B_c^\pm \to \psi(2S) \pi^\pm$ 
to $B_c^\pm \to J/\psi \pi^\pm$ branching ratios by 
correcting for the relative reconstruction 
efficiencies and found the value of 
\bea
\frac{B_c^\pm\to\psi(2S)\pi^\pm}{B_c^\pm \to J/\psi \pi^\pm} &=& 0.250 \pm 0.068(stat)\pm 0.014(syst)\pm 0.006 (\mathcal{B}),
\eea
where 
the third uncertainty is due to the uncertainties on the branching ratios of $J/\psi$ and 
$\psi(2S)$ in dimuons~\cite{pdg}.
The dominant source of systematic uncertainty is the one 
associated with the BDT selection, which amounts to 4.5\%.
The results are in good agreement with the theoretical predictions 
of 0.18 made by the relativistic quark model~\cite{bcRatioTh}.

\subsection{$B_c^\pm\to J/\psi D_s^{(*)\pm}$ observation}
Using the full dataset of $\simeq 3 \fbinv$ collected until 2012, we have 
observed the decays $B_c^\pm\to J/\psi D_s^{(*)\pm}$ for the
first time~\cite{bcjpsids}.
The $\jpsi$ is reconstructed in its dimuon decay, 
while the $D_s^{\pm}$ is reconstructed through its decay 
into $\phi\pi$, followed by $\phi\to K^\pm K^\mp$.
The decay  $B_c^\pm\to J/\psi D_s^{*}$ appears 
in the $\jpsi D_s$ invariant mass  as a satellite structure 
at smaller mass.
The number of signal events
for the two decays is determined by a fit to the 
$D_s^{\pm(*)} \pi$ invariant mass distribution,
shown in Fig.~\ref{fig:Fig1} (right), 
using a double Crystal Ball for the $D_s$ 
signal and the shapes obtained from the Monte Carlo
of the distributions due to the $\mathcal{A}^{\pm\pm},{A}^{00}$
different amplitudes for the $D_s^*$.
Using the $B_c^\pm\to\jpsi\pi^\pm$ as normalisation channel
we can measure the ratio of branching ratios 
\bea
\frac{B_c^\pm\to\jpsi D_s^\pm}{B_c^\pm \to J/\psi \pi^\pm} &=& 2.90 \pm 0.57(stat)\pm 0.24(syst),\\
\frac{B_c^\pm\to\jpsi D_s^{*\pm}}{B_c^\pm \to J/\psi D_s^\pm} &=& 2.37 \pm 0.56(stat)\pm 0.10(syst),
\eea
where the dominant systematic is the one associated with the 
knowledge of the branching ratio of $D_s^\pm\to\phi(\to K^\pm K^\mp)\pi^\pm$~\cite{pdg}.
These results are in good agreement with the simple 
factorisation approach but generally disagree with the 
other models~\cite{bcjpsids}.
Given the small Q value associated with this decay and 
the precise knowledge of the D meson mass differences~\cite{dmass}
it is possible to obtain a precise measurement of the 
$B_c$ mass, which is found to be 
$m(B_c^\pm) = 6276.26 \pm 1.44(stat) \pm 0.28(syst)~\MeVCSq$,
in excellent agreement with the previous 
LHCb result~\cite{bc3} and with the world average~\cite{pdg}.

\section{B hadron masses}\label{sec:bmasses}
At LHCb we reconstructed three of the 16 b-baryons predicted ground states,
namely the $\Lambda_b^0, \Xi_b^-$ and the $\Omega_b^-$ in their 
decays $\jpsi\Lambda^0, \jpsi\Xi^-$ and $\jpsi\Omega^-$
respectively.
Using a minimal set of selections in 1~\fbinv of data
we measured the masses~\cite{bmasses} of the above mentioned baryons, 
finding the values of 
\bea
m(\Lambda_b^0) &=& 5619.53 \pm 0.13 (stat) \pm 0.45 (syst) \MeVCSq,\\
m(\Xi_b^0)     &=& 5795.8 \pm 0.9 (stat) \pm 0.4 (syst) \MeVCSq,\\
m(\Omega_b^0)  &=& 6046.0 \pm 2.2 (stat) \pm 0.4 (syst) \MeVCSq,
\eea
with the dominant systematic uncertainty coming from the knowledge of 
the momentum scale.
These results are in agreement with the previous measurements
and with the world average~\cite{pdg}.


\section{X(3872) quantum numbers}\label{sec:xjpc}
\begin{figure}
\begin{minipage}{0.45\linewidth}
\centerline{\includegraphics[width=0.9\linewidth]{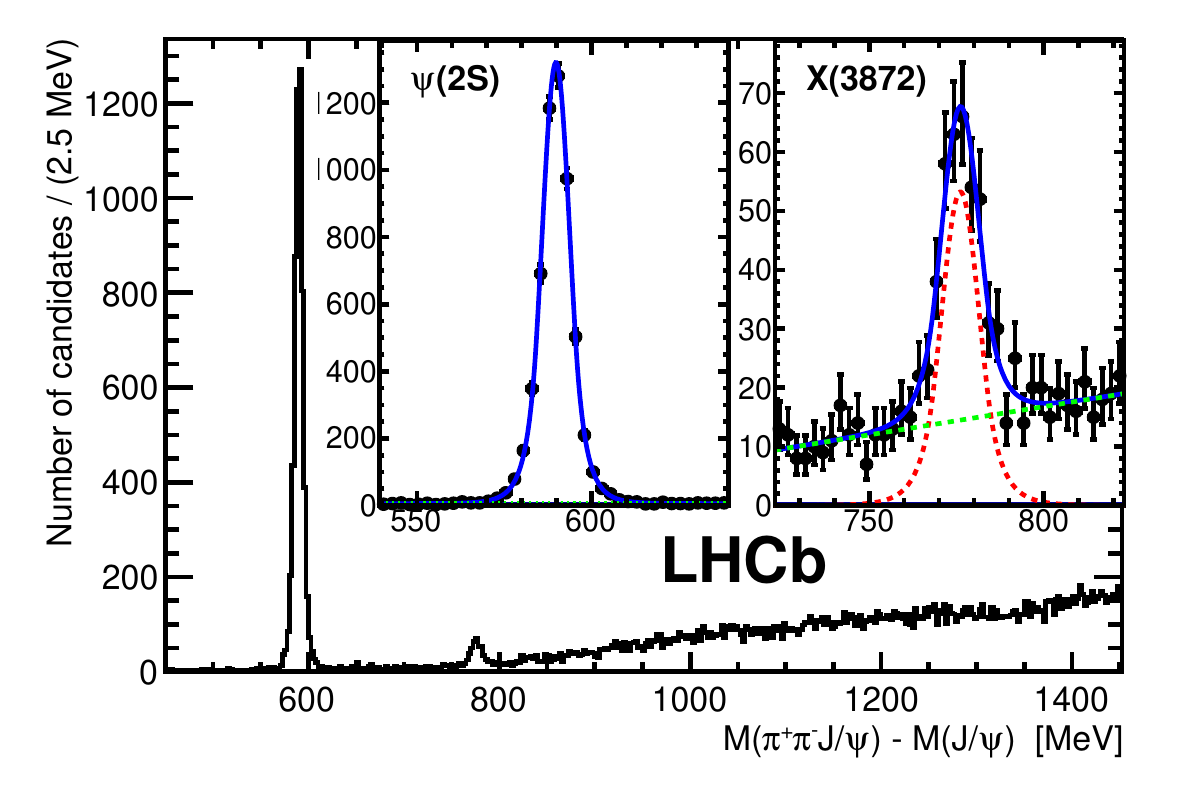}}
\end{minipage}
\hfill
\begin{minipage}{0.45\linewidth}
\centerline{\includegraphics[width=0.9\linewidth]{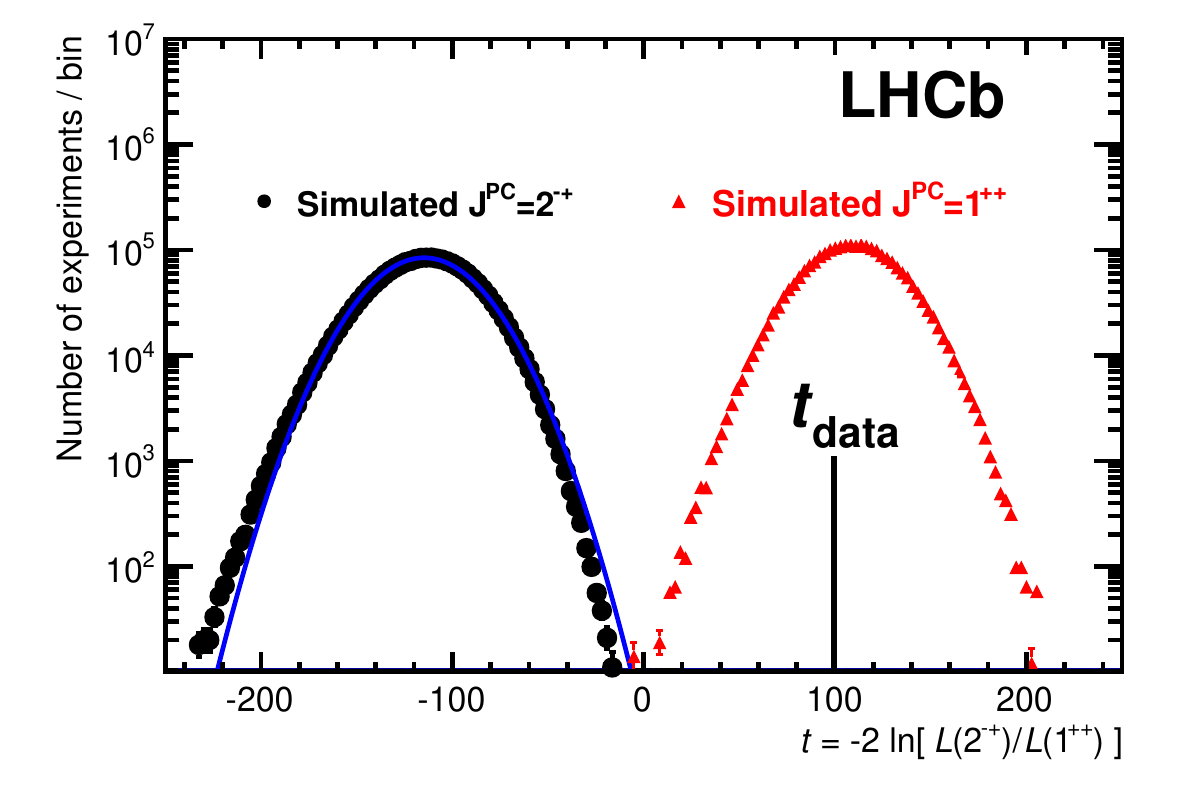}}
\end{minipage}
\hfill
\caption{{\it Left:}
Distribution of $m(\jpsi\pi^\pm\pi^\mp)-m(\jpsi)$ for
$B^\pm\to\jpsi\pi^\pm\pi^\mp K^\pm$ candidates. The results of the fit 
around the $\psi(2S)$ and $X$ masses are shown in the
inserts. The solid blue, dashed red, and dotted green lines represent the
total, signal, and background component, respectively.
{\it Right:}
Distribution of the test statistic 
for the simulated experiments with
$J^{PC}= 2^{-+}$ (black circles) and with
$J^{PC}= 1^{++}$ (red triangles). A Gaussian fit 
to the two distributions is overlaid (blue solid line). 
The value of the test statistic for the data, ``$t_{data}$'',
is shown by the solid vertical line.}
\label{fig:Fig2}
\end{figure}

The X(3872) (called X in the rest of the paper)
has been the first exotic state to be discovered~\cite{xbelle}
and by far the most abundant.
Its mass just above the $DD^*$ threshold still intrigues 
theorists and experimentalists who wonder about its real nature.
After measuring the mass and the 
production cross section~\cite{xlhcb},
we have measured the X quantum numbers~\cite{xjpc}, resolving 
the ambiguity observed by Belle between $1^{++}$ and $2^{-+}$
in favour of the former.
In 1.1~\fbinv of data we have performed a five dimensional analysis
of 313$\pm$26  $B^+\to X K^+$ decays, with $X\to\jpsi\pi^\pm\pi^\mp$ and $\jpsi\to\mu^\pm\mu^\mp$.
The selection is optimised on the $B^\pm\to\psi(2S)K^\pm$ similar channel, 
and the signal is determined through a fit to the data 
using a Crystal Ball function for the signal and a 
linear function for the background, as shown in Fig.~\ref{fig:Fig2}, left.
A likelihood ratio test is performed to discriminate 
between the two quantum numbers hypotheses, which shows that the $1^{++}$ 
option is favoured and the $2^{-+}$ option is rejected at 8.4$\sigma$
(see Fig.~\ref{fig:Fig2}, right).

\section{Conclusions}
LHCb has a flourishing program in spectroscopy which is 
getting more and more interesting as more and more data are collected.
Important results have already been achieved especially 
in exotic spectroscopy and more are expected in the 
near future.

%

\end{document}